\documentclass[11pt]{article}
\pdfoutput=1
\usepackage{graphicx}
\usepackage{hyperref}
\usepackage{epsfig}
\usepackage{amsmath}
\usepackage{amsthm}
\usepackage{amssymb}
\usepackage{multirow}
\usepackage{color}
\usepackage[nosort]{cite}
\usepackage{hyperref}
\usepackage[fontsize=11.45pt]{scrextend}
\usepackage{braket,mathrsfs,slashed}
\usepackage{float}
\usepackage{setspace}
\usepackage[caption=false]{subfig}
\newlength{\abstractwidth}
\newcommand{\be}{\begin{equation}}
\newcommand{\ee}{\end{equation}}

\renewcommand{\title}[1]{\vbox{\center\bf{\Large{#1}}}\vspace{5mm}}
\renewcommand{\author}[1]{\vbox{\center#1}\vspace{5mm}}
\newcommand{\address}[1]{\vbox{\center\em#1}}

\usepackage{dsfont}

\usepackage[utf8]{inputenc}
\usepackage{rotating}
\usepackage{dsfont}
\usepackage[paper=a4paper,margin=1.5cm]{geometry}
\usepackage[utf8]{inputenc}
\usepackage{rotating}
\usepackage{soul}

\usepackage{mathrsfs} 
\usepackage{bbm}
\usepackage{etoolbox} 
\usepackage{multirow}

\renewcommand\[{\begin{equation}}
\renewcommand\]{\end{equation}}

\newcommand{\ba}{\begin{eqnarray}}
\newcommand{\ea}{\end{eqnarray}}

\hypersetup{pdfstartview=FitV,colorlinks=true,linkcolor=midblue,citecolor=midblue,filecolor=midblue,urlcolor=midblue}

\definecolor{midblue}{rgb}{0,0,0.5}

\begin{document}
	
		\newgeometry{top=3.1cm,bottom=3.1cm,right=2.4cm,left=2.4cm}
		
	\begin{titlepage}
	\begin{center}
		\hfill \\
		\vskip 0.5cm

		\title{Non-local positivity bounds: \\[2mm]
        islands in Terra Incognita}

			\author{\large Luca Buoninfante$^{a,\,\star}$, Long-Qi Shao$^{b,c,d\,\dagger}$, Anna Tokareva$^{b,e,f,\,\ddagger}$ }
			
			\address{$^a$High Energy Physics Department, Institute for Mathematics, Astrophysics,\\
			and Particle Physics, Radboud University, Nijmegen, The Netherlands\\[1.5mm]
				$^b$School of Fundamental Physics and Mathematical Sciences, \\Hangzhou Institute for Advanced Study, UCAS, Hangzhou 310024, China\\[1.5mm]
                $^c$Institute of Theoretical Physics, Chinese Academy of Sciences, Beijing 100190, China\\[1.5mm]
                $^d$University of Chinese Academy of Sciences, Beijing 100049, China\\[1.5mm]
                $^e$International Centre for Theoretical Physics Asia-Pacific, Beijing/Hangzhou, China\\[1.5mm]
                $^f$Department of Physics, Blackett Laboratory, Imperial College London, SW7 2AZ London, UK
                }
				\vspace{.3cm}

		\end{center}

\vspace{0.15cm}

\begin{abstract}
The requirements of unitarity and causality lead to significant constraints on the Wilson coefficients of an EFT expansion, known as positivity bounds. Their standard derivation relies on the crucial assumption of polynomial boundedness on the growth of scattering amplitudes in the complex energy plane, which is a property satisfied by local QFTs, and by weakly coupled string theory in the Regge regime. 
The scope of this work is to clarify the role of locality by deriving generalized positivity bounds under the assumption of exponential boundedness, typical of non-local QFTs where the Froissart-Martin bound is usually not satisfied.
Using appropriately modified dispersion relations, we derive new constraints and find regions in the EFT parameter space that do not admit a local UV completion. Furthermore, we show that there exist EFTs that satisfy IR causality and at the same time can admit a non-local UV completion, provided that the energy scale of non-locality is of the same order or larger than the EFT cutoff. Finally, we provide an explicit example of an exponentially bounded amplitude that satisfies partial-wave unitarity and asymptotic causality.
\end{abstract}
\vspace{4cm}
\noindent\rule{6.5cm}{0.4pt}\\
$\,^\star$
\href{mailto:luca.buoninfante@ru.nl}{luca.buoninfante@ru.nl}\\	
$\,^\dagger$ \href{mailto:shaolongqi22@mails.ucas.ac.cn}{shaolongqi22@mails.ucas.ac.cn}\\
$\,^\ddagger$ \href{mailto:tokareva@ucas.ac.cn}{tokareva@ucas.ac.cn}

\end{titlepage}


\baselineskip=17.63pt



\newpage

\textit{\textbf{Introduction.---}} 
To constrain the parameter space of viable effective field theories (EFTs) it is often not necessary to know the precise form of the UV completion.
In many cases it is sufficient to assume certain physical properties in the UV regime, which can be translated into mathematical requirements on the $S$-matrix, from which restrictions on low-energy physics can then be derived~\cite{Pham:1985cr,Pennington:1994kc, Adams:2006sv,Komargodski:2011vj,Arkani-Hamed:2020blm,Caron-Huot:2020cmc,Bellazzini:2020cot,Remmen:2019cyz,Bellazzini:2019xts,Herrero-Valea:2019hde,Bellazzini:2017fep,deRham:2017avq,deRham:2017zjm,deRham:2017imi,Wang:2020jxr,Tolley:2020gtv,Alberte:2020bdz, Tokuda:2020mlf,Li:2021lpe,Caron-Huot:2021rmr,Du:2021byy,Bern:2021ppb,Li:2022rag, Caron-Huot:2022ugt,Saraswat:2016eaz, Arkani-Hamed:2021ajd,Herrero-Valea:2020wxz,Guerrieri:2021ivu,Henriksson:2021ymi,Henriksson:2022oeu,EliasMiro:2022xaa,Bellazzini:2021oaj, Herrero-Valea:2022lfd,Hong:2023zgm,Chiang:2022jep,Huang:2020nqy,Noumi:2021uuv,Chiang:2022ltp, Xu:2023lpq, Chen:2023bhu,Noumi:2022wwf, Hong:2024fbl,Bern:2022yes,Ma:2023vgc, DeAngelis:2023bmd,Aoki:2023khq,Xu:2024iao,EliasMiro:2023fqi,McPeak:2023wmq,Riembau:2022yse,Caron-Huot:2024tsk,Caron-Huot:2024lbf,Wan:2024eto} (see also~\cite{deRham:2022hpx} for a recent review). For example, standard requirements that are usually assumed are those of Lorentz invariance, unitarity and causality. These properties are encoded in a set of mathematical conditions, such as positivity of the imaginary part of the partial waves and analyticity of the $2\rightarrow 2$ scattering amplitude in the complex energy plane~\cite{Bogolyubov:1959bfo, Toll:1956cya, Bremermann:1958zz, Eden:1966dnq}. 

A further assumption that is typically made on the UV completion is that of \textit{locality}, which states that the growth of the scattering amplitude is \textit{polynomially bounded} in the complex energy plane.
For local gapped theories it can be proven that the amplitude for fixed physical momentum transfer is polynomially bounded by $|s|^2$ in the whole complex plane; this is known as Froissart-Martin bound\footnote{It is worth to mention that Froissart~\cite{Froissart:1961ux} and Martin~\cite{Martin:1962rt} initially showed that the amplitude was bounded by $|s|$ in the forward limit $t=0.$ Subsequently, Jin and Martin~\cite{Jin:1964zz} showed that the amplitude for generic physical momentum transfer $t\leq 0$ is bounded by $|s|^2.$ In any case, in this work we refer to the latter as Froissart-Martin bound.}~\cite{Froissart:1961ux,Martin:1962rt,Jin:1964zz}.
Although for theories with massless states no rigorous general proof of this bound exists, it is common to assume that the growth of the amplitude for fixed physical $t$ is bounded by $|s|^2$ in the whole complex plane:\footnote{This can be verified at least for graviton-mediated amplitudes in $d>4$, using the method of eikonal resummation~\cite{Haring:2022cyf,Haring:2024wyz}.}
\begin{equation}    \label{eq:martin-froissart}
    |A(s,t\leq 0)|<B|s|^2\,, \qquad  |s|\rightarrow \infty\,,
\end{equation}
where $B$ is a subdominant factor that may depend on $s$.

To clarify the role of locality, it is instructive to investigate what happens when we allow for a non-polynomial growth of the scattering amplitude, i.e. for violations of the Froissart-Martin bound~\eqref{eq:martin-froissart}. In particular, we are going to consider amplitudes that are exponentially bounded in the complex $s$ plane for fixed physical $t$:
\begin{equation}    \label{eq:exp-bounded}
    |A(s,t\leq 0)|<C|s|^2 e^{\alpha |s|^{2g}}\,,\qquad  |s|\rightarrow \infty\,,
\end{equation}
where $\alpha$ and $g$ are two positive constant parameters, and $C$ a subdominant factor that may depend on $s$.

QFTs in which scattering amplitudes are characterized by this type of exponential behavior could be classified according to Jaffe's criterion~\cite{Jaffe:1966an,Jaffe:1967nb,Keltner:2015xda, Tokuda:2019nqb,Buoninfante:2023dyd}: if $g< 1/4$ the corresponding QFT is said to be \textit{localizable}, whereas if $g\geq 1/4$ the QFT is said to be \textit{non-localizable}. When $g=0$ we recover the standard local QFT description with polynomially bounded amplitudes. The name non-localizability is related to the fact that the fields must be delocalized through smeared test functions in order to make the position-space correlators (such as the Wightman function) finite and well-defined. In what follows we use the words local (non-local) and localizable (non-localizable) interchangeably.

The aim of this Letter is to understand what type of constraints can be derived on the Wilson coefficients of the low-energy EFT expansion if the UV completion is assumed to be non-localizable according to~\eqref{eq:exp-bounded}, and study their physical implications. 
We work in four spacetime dimensions, and adopt the mostly plus metric signature $(-+++)$ and the Natural units system $(\hbar=1=c).$ 

\textit{\textbf{Dispersion relations.---}} The central object of our investigation is the elastic $2\rightarrow 2$ scattering amplitude $A(s,t,u)$ in some gapless scalar theory, where $s=-(p_1+p_2)^2,$ $t=-(p_1-p_3)^2,$ $u=-(p_1-p_4)^2$ are the Mandelstam variables satisfying the relation $s+t+u=0$; $p_1,$ $p_2$ are the ingoing momenta and $p_3,$ $p_4$ are the outgoing momenta.  In particular, $s$ is the centre-of-mass energy squared, while $t$ is the momentum transfer squared. The scattering amplitude can be chosen to be a function of two Mandelstam variables only, e.g. $A(s,t,u)=A(s,t)$.

We assume that the low-energy behavior of the amplitude can be captured by the following EFT expansion: 
\begin{equation} \label{A-eft-expansion}
\begin{aligned}
    A_{\rm EFT}(s,t)=&\,g_2(s^2+t^2+u^2)+g_3 s t u+g_4 (s^2+t^2+u^2)^2 \\
    &+g_5(s t u)(s^2+t^2+u^2) +g_6 (s^2+t^2+u^2)^3+\cdots \,,
    \end{aligned}
\end{equation}
which is manifestly crossing symmetric; the couplings $g_n$ are called Wilson coefficients.

\begin{figure}[t!]
    \centering
    \includegraphics[scale=0.4]{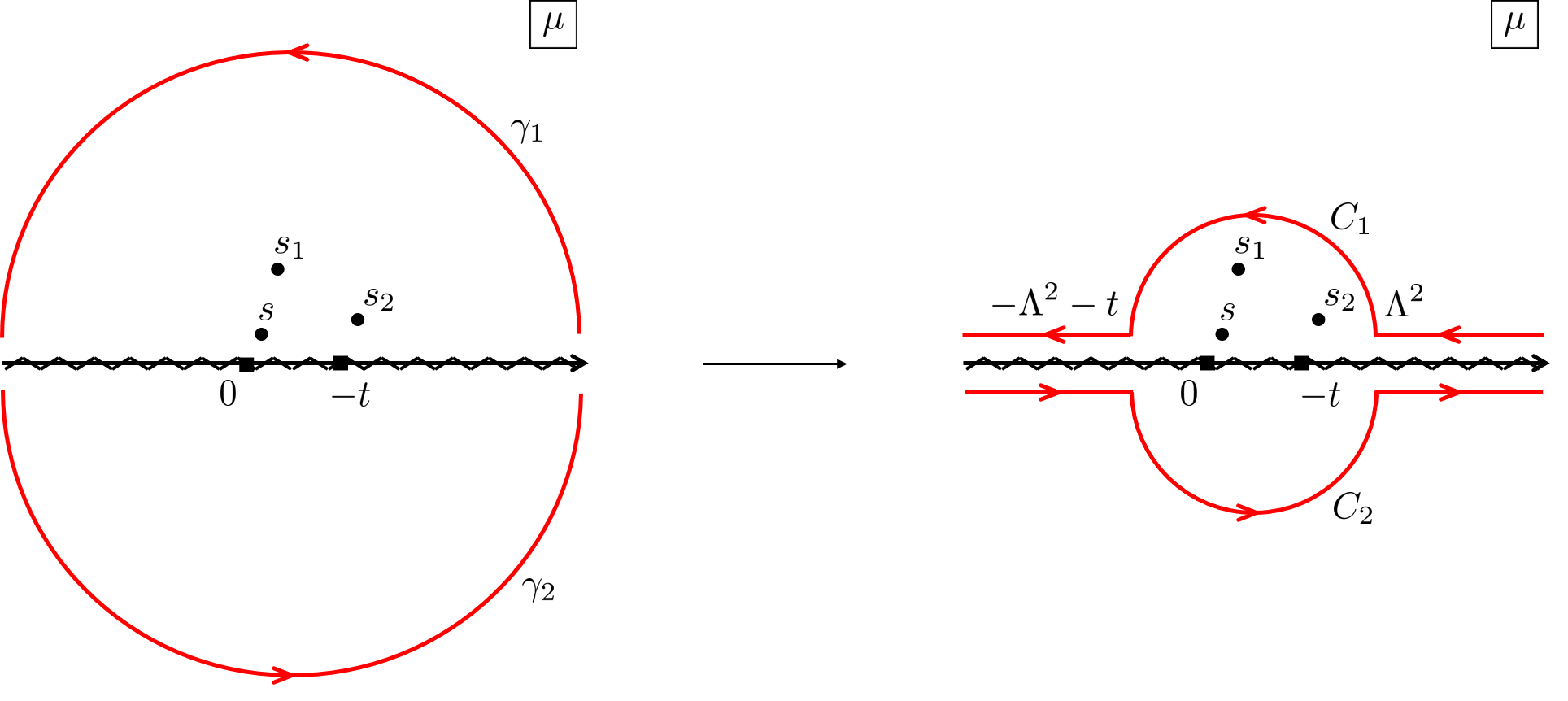}
    \caption{Integration contour in the complex $\mu$ plane used to derive the twice-subtracted dispersion relations in Eqs.~\eqref{form:dispersion-zero} and~\eqref{form:dispersion2} (with $s_1=s_2=0$). The initial contour $\gamma_1\cup \gamma_2$ on the left is taken of infinite radius. It is then deformed into the contour on the right, where the circle $C_1\cup C_2$ is centered at $\mu=-t/2.$ The boxes denote the two branch points of the $s$-channel (from $\mu=0$ to $+\infty$) and $u$-channel (from $\mu=-t$ to $-\infty$) branch cuts, respectively.  The point $s$, $s_1$ and $s_2$ are the poles in Eq.~\eqref{integral-zero}.}
    \label{fig:contour}
\end{figure}

In the past literature~\cite{Adams:2006sv,deRham:2022hpx}, the Wilson coefficients have been constrained assuming that the UV completion is local, i.e. the amplitude is polynomially bounded in all complex directions in the $s$ and $t$ planes. In addition to these ingredients, the assumption of analyticity of the amplitude allows to derive various dispersion relations at fixed $t\leq 0$, connecting the Wilson coefficients in $A_{\rm EFT}$ with positive contributions coming from the discontinuities along the branch cuts of the amplitude. Indeed, by choosing the contour $\gamma_1\cup \gamma_2$ in Fig.~\ref{fig:contour}, we can write~\cite{Caron-Huot:2020cmc,Caron-Huot:2022ugt}
\begin{equation}\label{integral-zero}
\int_{\gamma_1\cup\gamma_2}\frac{{\rm d}\mu}{2\pi i} \frac{A(\mu,t)}{(\mu-s_1)(\mu-s_2)(\mu-s)}=0\,,
\end{equation}
where $s_1$ and $s_2$ are two subtraction points. Subsequently, by deforming the integration contour as shown in Fig.~\ref{fig:contour}, Eq.~\eqref{integral-zero} can be recast as
\begin{equation}\label{integral-zero-2}
\begin{aligned}
\int_{C_1\cup C_2}\frac{{\rm d}\mu}{2\pi i} \frac{A(\mu,t)}{(\mu-s_1)(\mu-s_2)(\mu-s)}
=&\,\int_{\Lambda^2}^\infty \frac{{\rm d}\mu}{\pi} \frac{{\rm Disc}_s A(\mu,t)}{(\mu-s_1)(\mu-s_2)(\mu-s)} \\
&+ \int_{\Lambda^2}^\infty \frac{{\rm d}\mu}{\pi} \frac{{\rm Disc}_s A(\mu,t)}{(\mu-u_1)(\mu-u_2)(\mu-u)}  \,,
\end{aligned}
\end{equation}
where $ {\rm Disc}_sA(\mu,t)=\lim_{\epsilon\rightarrow0^+} \frac{1}{2i}[A(\mu+i\epsilon,t)-A(\mu-i\epsilon,t)];$
we have used crossing symmetry to write ${\rm Disc}_sA(-\mu-t,t)=-{\rm Disc}_sA(\mu,t)$ and have defined the similar subtraction points in terms of the Mandelstam variable $u,$ i.e. $u_1=-s_1-t$ and $u_2=-s_2-t.$ The scale $\Lambda$ represents the EFT cutoff.

By computing the integral on the left-hand-side of Eq.~\eqref{integral-zero-2} through the residue theorem, we can obtain the twice-subtracted dispersion relation. For simplicity we choose $s_1=s_2=0,$ thus we can write
\begin{equation}
\begin{aligned}\label{form:dispersion-zero}
A_{\rm EFT}(s,t)=s^2\int_{\Lambda^2}^\infty \frac{{\rm d}\mu}{\pi} \frac{{\rm Disc}_s A(\mu,t)}{\mu^2(\mu-s)}+ s^2\int_{\Lambda^2}^\infty \frac{{\rm d}\mu}{\pi} \frac{{\rm Disc}_s A(\mu,t)}{(\mu+t)^2(\mu-u)}  + a(t) + s\, b(t)\,,
\end{aligned}
\end{equation}
where
\begin{equation}\label{subt-terms-zero}
a(t)= A_{\rm EFT}(0,t)\,,\qquad b(t)= \left.\frac{\partial A_{\rm EFT}(s,t)}{\partial s}\right|_{s=0}\,,
\end{equation}
are the two subtraction terms. Note that, we wrote $A_{\rm EFT}(s,t)$ because the left-hand-side of Eq.~\eqref{form:dispersion-zero} satisfies the equation only for $s< \Lambda^2.$ This aspect is what allows us to make the connection with the low-energy EFT expansion~\eqref{A-eft-expansion}. In fact, more precisely, the left-hand-side of Eq.~\eqref{form:dispersion-zero} should read $A(s,t)|_{s<\Lambda^2}\simeq A_{\rm EFT}(s,t).$

If the assumption of polynomial boundedness is given up, the integral over $\gamma_1\cup \gamma_2$ in Eq.~\eqref{integral-zero} could blow up.\footnote{It is worth to mention that for localizable QFTs, in which the amplitude is bounded as~\eqref{eq:exp-bounded} with $g<1/4,$ it can actually be shown that the amplitude cannot grow faster than a polynomial in the complex $s$ plane~\cite{Tokuda:2019nqb}. In this Letter, instead, we are interested in non-localizable QFTs with $g>1/4$ (in particular $g=1$), for which a similar result does not necessarily apply. Therefore, modified dispersion relations that take into account a non-localizable  exponential growth are needed in our case.} This means that we need to implement a modified dispersion relation. Let us start considering the forward limit $t=0,$ and later on we will address the case of non-vanishing physical momentum transfer that will require a more careful treatment.

When $t=0$, we can assume that the amplitude is bounded as $A(s,t=0)< s^2 \exp{(-\alpha s^{2g})}$ for any $s\in \mathbb{C},$ up to subdominant factors. This means that along the real axis the amplitude is exponentially suppressed in the forward limit, and can grow exponentially along some direction in the complex $s$ plane. To write down a dispersion relation for this type of amplitude, we can adopt the following strategy. We multiply the amplitude by $\exp{(\alpha s^{2g})}$ and define $A'(s,t)\equiv \exp{(\alpha s^{2g})}A(s,t)$, which will be bounded by $|s|^2$ in the whole complex $s$ plane. In other words, we compensate for the essential singularity at infinity, so that we can write a dispersion relation for $A'(s,t=0)$ that is analogous to Eq.~\eqref{form:dispersion-zero}:
\begin{equation}  
\begin{aligned}\label{form:dispersion2}
    \exp{(\alpha s^{2g})}A(s,0)=s^2\int_{\Lambda^2}^{\infty} \frac{{\rm d}\mu}{\pi} \frac{\rm{Disc}_s A'(\mu,0)}{\mu^2(\mu-s)}
    +s^2\int_{\Lambda^2}^{\infty} \frac{{\rm d}\mu}{\pi} \frac{\rm{Disc}_s A'(\mu,0)}{\mu^2(\mu+s)}+a'(0)+sb'(0)\,,
\end{aligned}
\end{equation}
where $a'(0)$ and $b'(0)$ have the same form as those in Eq.~\eqref{subt-terms-zero}, but with $A$ replaced by $A'$. However, since our amplitude satisfies $A_{\rm EFT}(0,0)=0$ and $\partial_sA_{\rm EFT}(s,0)|_{s=0}=0,$ the two subtraction terms actually vanish in the forward limit, i.e. $a'(0)=0=b'(0).$

\textit{\textbf{Positivity bounds for $\boldsymbol{t=0}$.---}} Let us now derive the positivity bounds in the forward limit. Hereafter we study the non-localizable case with $g=1$.

If we define the positive quantity
\begin{equation}
   \left < \frac{1}{\mu^m} \right>\equiv \int_{\Lambda^2}^{\infty} \frac{{\rm d}\mu}{\pi} {\rm Disc}_s A'(\mu,0)\frac{1}{\mu^m}\,,
\end{equation}
we can write
\begin{equation}
\begin{aligned}
   \left < 1/\mu^3 \right> =\,g_2\,,\quad
   \left <1/\mu^5 \right>=&\,2g_4+\alpha g_2\,,\\
  \left <1/\mu^7 \right>=\,4g_6+2\alpha g_4+\alpha^2 g_2/2\,,\quad
  \left <1/\mu^9 \right>=&\,g_8+4\alpha g_6+ \alpha^2 g_4+\alpha^3g_2/6\,,
   \end{aligned}
\end{equation}
and analogous relations can be obtained for higher-order Wilson coefficients. In addition to the positivity relations above, we also have the ``relaxing'' inequality $\left < 1/\mu^n \right> < \frac{1}{\Lambda^2}\left < 1/\mu^{n-1} \right>$, and the two Cauchy-Schwartz inequalities $\left < 1/\mu^{5} \right>^2<\left < 1/\mu^{3} \right>\left < 1/\mu^{7} \right>$ and $\left < 1/\mu^{7} \right>^2<\left < 1/\mu^{5} \right>\left < 1/\mu^{9} \right>$. This set of inequalities provides,
\begin{equation}
\begin{aligned}
   0<2\tilde{g}_4+\tilde{\alpha} &< 1, \qquad 0<4\tilde{g}_6+2\tilde{\alpha} \tilde{g}_4+\tilde{\alpha}^2/2 <2\tilde{g}_4+\tilde{\alpha}\,,\\ &4\tilde{g}_6+2\tilde{\alpha} \tilde{g}_4+\tilde{\alpha}^2/2 \geq(2\tilde{g}_4+\tilde{\alpha})^2\,,  \label{form:NLg4}
    \end{aligned}
\end{equation}
where we have defined the dimensionless quantities $\tilde{\alpha}\equiv \alpha \Lambda^4,$ $\tilde{g}_4\equiv g_4\Lambda^4/g_2$, $\tilde{g}_6\equiv g_6\Lambda^8/g_2$,  and so on. In the $\alpha \rightarrow 0$ limit, we recover the well-known local case~\cite{Bellazzini:2020cot}. 
\begin{figure*}[t!]%
    \centering
    \subfloat[\centering ]{\includegraphics[width=0.4\textwidth]{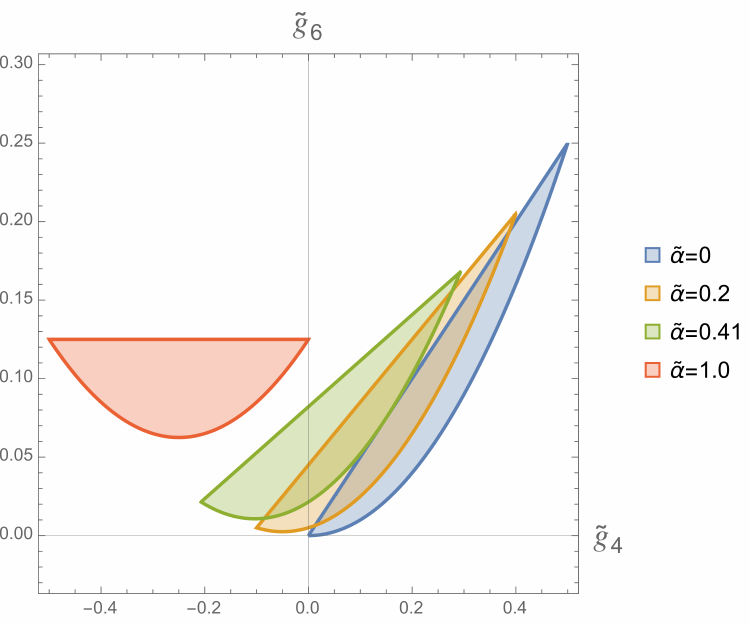}\label{fig.1.a}}%
    \qquad\qquad
    \subfloat[\centering ]{\includegraphics[width=0.4\textwidth]{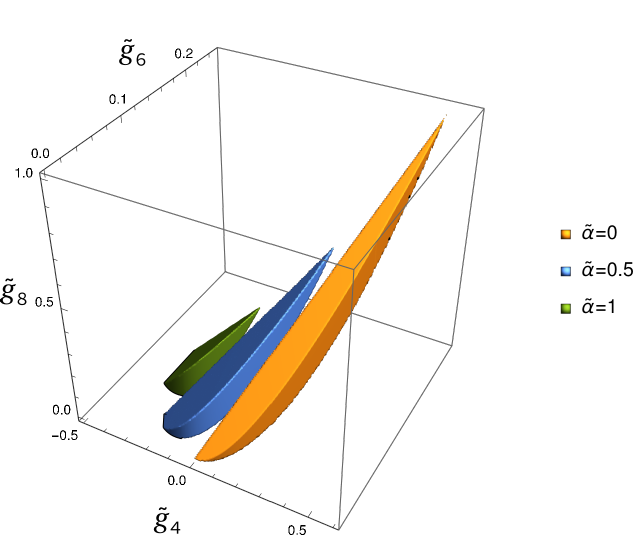}\label{fig.1.b}}%
    \caption{(a) Allowed regions in the $\tilde{g}_4$-$\tilde{g}_6$ plane when $\tilde{\alpha}=0,0.2,0.41,1$. (b) Allowed regions in the three-dimensional parameter space spanned by $\tilde{g}_4$, $\tilde{g}_6$ and $\tilde{g}_8$ when $\tilde{\alpha}=0,0.5,1$.} 
    \label{fig:g4g6g8}%
\end{figure*}

The regions in the $\tilde{g}_4$-$\tilde{g}_6$ plane that are allowed by the constraints in~\eqref{form:NLg4} are shown in Fig.~\ref{fig.1.a} for different values of $\tilde{\alpha}$. We can track how the allowed region changes as a function of $\tilde{\alpha}$, going from $\tilde{\alpha}=0$ to $\tilde{\alpha}=1.0$. In particular, we found that the regions corresponding to $\tilde{\alpha}\geq \sqrt{2}-1\approx0.41$ have no overlap with the local case $\tilde{\alpha}=0$. We also verified that using two Cauchy-Schwartz inequalities to derive the bounds on $\tilde{g}_4$ and $\tilde{g}_6$ is already sufficient to obtain the most optimal constraints, in agreement with the analyses performed in local cases~\cite{Bellazzini:2020cot,2105.02862}.
In Fig.~\ref{fig.1.b} we plotted the allowed regions in the three-dimensional parameter space spanned by $\tilde{g}_4$, $\tilde{g}_6$, $\tilde{g}_8$. 

\textit{\textbf{Positivity bounds for $\boldsymbol{t<0}$.---}} The above derivation of positivity bounds does not give us any information about the couplings in the EFT amplitude~\eqref{A-eft-expansion} that disappear when $t=0$, such as $g_3$. Constructing the dispersion relation for $t<0$ requires more details about the structure of the UV amplitude. 
We assume the following crossing-symmetric form, which is consistent with partial-wave unitary (see the Supplemental Material):
\begin{equation}
\label{UV crossing}
    A(s,t,u)=F(t,u)e^{-\alpha s^{2}}+F(s,u)e^{-\alpha t^{2}}+F(s,t)e^{-\alpha u^{2}}\,,
\end{equation}
where $F(s,t)$ is bounded by $s^2$ in the entire complex plane. We can then write the $s\leftrightarrow u$ crossing-symmetric representation as
\begin{equation}
\label{su}
\begin{aligned}
    A_{\rm{EFT}}(s,t)=\,
    e^{-\alpha s^{2}}s^2\int_{\Lambda^2}^{\infty} \frac{{\rm d}\mu}{\pi} \frac{{\rm Disc}_s A'(\mu,t)}{\mu^2(\mu-s)}+e^{-\alpha u^{2}}u^2\int_{\Lambda^2}^{\infty} \frac{{\rm d}\mu}{\pi} \frac{{\rm Disc}_s A'(\mu,t)}{\mu^2(\mu-u)}+a(t)\,,
    \end{aligned}
\end{equation}
where $A'(\mu,t)=A(\mu,t)e^{\alpha \mu^{2}}$. The expression for $a(t)$ can be derived by imposing full crossing symmetry and we have $a(0)=0$. Note that, there is no subtraction term $s\,b(t)$, otherwise the $s\leftrightarrow u$ crossing symmetry would be broken.

\begin{figure*}%
\label{IRcausality}
    \centering
    \subfloat[\centering ]{\includegraphics[width=0.4\textwidth]{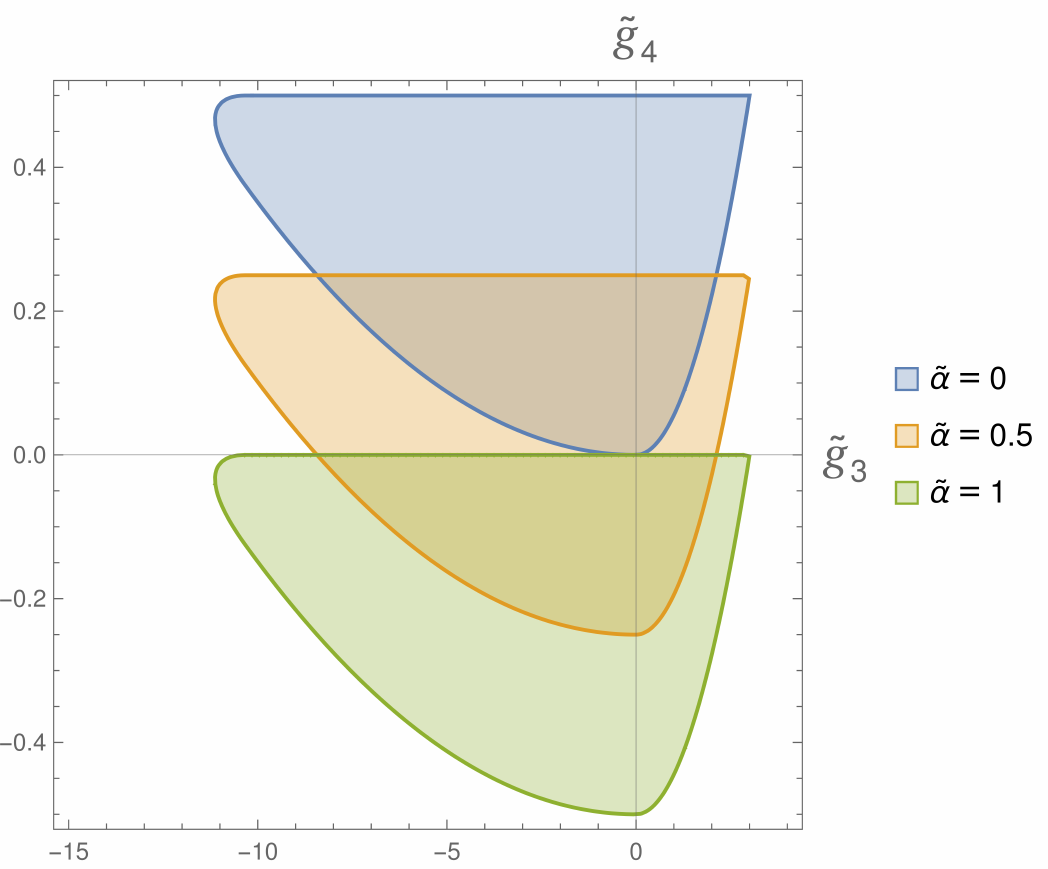}\label{fig3.a}}%
    \qquad\qquad
    \subfloat[\centering ]{\includegraphics[width=0.42\textwidth]{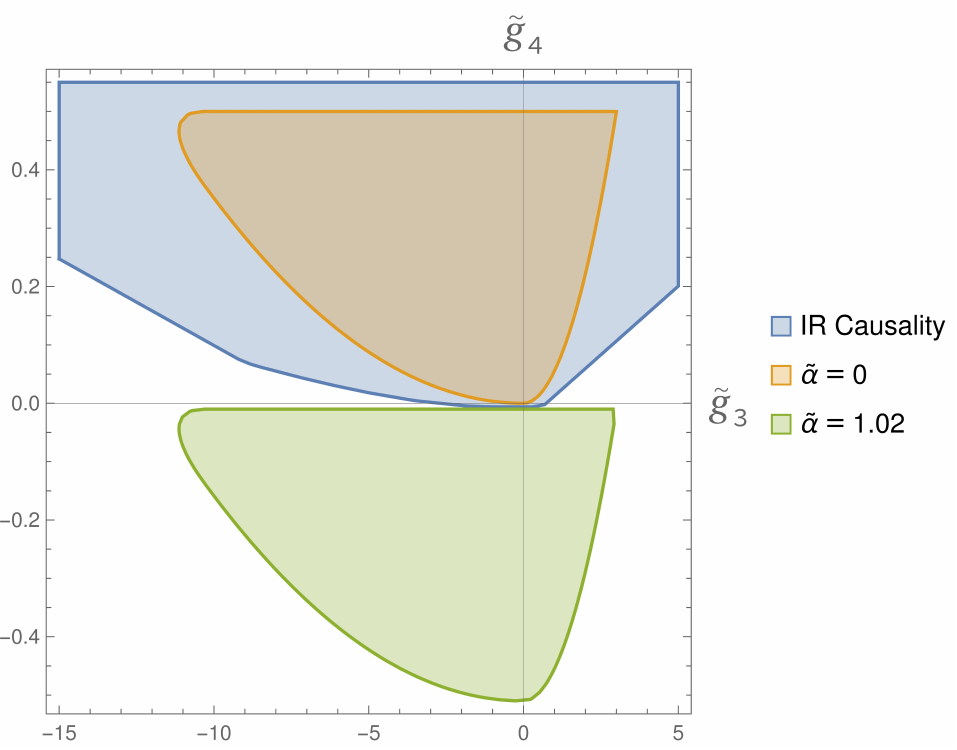}\label{fig3.b}}%
    \caption{(a) Allowed region in the $\tilde{g}_3$-$\tilde{g}_4$ plane for $\tilde{\alpha}=0,0.5,1$ from~\eqref{ineq} and~\eqref{ineq2}. (b) Allowed regions by IR causality (blue region)~\cite{CarrilloGonzalez:2022fwg} and  positivity bounds on $\tilde{g}_3$ and $\tilde{g}_4$ in  both local ($\tilde{\alpha}=0$) and non-local ($\tilde{\alpha}=1.02$) cases. For $\tilde{\alpha}<1.02$ there is an overlap between constraints from positivity bounds and IR causality, while larger values of $\tilde{\alpha}$ are in contrast with IR causality.} 
    \label{fig:g3h3f3slice}%
\end{figure*}

The state-of-the-art technique to derive the bound on $g_3$ is based on the theory of moments and is described in~\cite{2105.02862,2308.11692}. We briefly review the formalism in the Supplemental Material and also show that it does not depend on the particular definition of the brackets. In other words, up to additional $\alpha$-dependent terms in some of the relevant relations, the constraints in the $g_3$-$g_4$ plane in our non-local model can be derived by following the same steps as in the local case~\cite{2105.02862,2308.11692}. The result is given by the following two sets of inequalities:
\begin{eqnarray}
\label{ineq}
    \textbf{I:} && \left\lbrace \begin{array}{l}
    \displaystyle -\sqrt{\frac{427}{3}(2\tilde{g}_{4}+\tilde{\alpha})}<\tilde{g}_{3}<3\sqrt{2\tilde{g}_{4}+\tilde{\alpha}}\,,\\[1.5mm]
    \displaystyle -\frac{\tilde{\alpha}}{2}\leq\tilde{g}_{4}\leq\frac{961}{2562}-\frac{\tilde{\alpha}}{2}\,,
    \end{array}\right.\\[2mm]
    \textbf{II:} && \left\lbrace \begin{array}{l}
    \label{ineq2}
    \displaystyle -\frac{62}{6}(2\tilde{g}_{4}+\tilde{\alpha})-\sqrt{\frac{320}{9}\left[(2\tilde{g}_{4}+\tilde{\alpha})-(2\tilde{g}_{4}+\tilde{\alpha})^2\right]}<\tilde{g}_{3}<3\sqrt{2\tilde{g}_{4}+\tilde{\alpha}}\,,\\[1.5mm]
    \displaystyle \frac{961}{2562}-\frac{\tilde{\alpha}}{2}\leq\tilde{g}_{4}\leq\frac{1-\tilde{\alpha}}{2}\,.
    \end{array}\right.
\end{eqnarray}
In Fig.~\ref{fig3.a} we have shown the allowed regions in the $\tilde{g}_3$-$\tilde{g}_4$ plane given by the bounds~\eqref{ineq} and~\eqref{ineq2}, for different values of $\tilde{\alpha}$. The local constraint is just shifted down allowing for negative values of $g_4$, which are forbidden in the standard local case.

\textit{\textbf{IR causality bounds.---}} The requirement of causality for the low-energy theory imposes a set of bounds on the Wilson coefficients of the EFT expansion~\eqref{A-eft-expansion}, that must hold independently of the type of UV completion. Causality can be probed by considering a process in which a plane wave propagates on a classical background created by some local source~\cite{Camanho:2014apa,Camanho:2016opx,Bai:2016hui,Goon:2016une,Hinterbichler:2017qcl,Hinterbichler:2017qyt,AccettulliHuber:2020oou,Bellazzini:2021shn,Serra:2022pzl,Chen:2021bvg,deRham:2021bll,CarrilloGonzalez:2022fwg,CarrilloGonzalez:2023cbf, CarrilloGonzalez:2023emp}, and its validity demands the absence of superluminal propagation. 
However, the uncertainty principle prevents us from detecting very small time advances in the signal propagation, i.e. $|\Delta T| \lesssim 1/\omega,$ where $\omega$ is the energy of the propagating wave. Therefore, unobservable causality violations could still be physically allowed because, thanks to their quantum stochastic nature, they do not sum up into macroscopic time advances in the signal propagation~\cite{1512.04952,Serra:2022pzl,CarrilloGonzalez:2022fwg,2312.07651}. This means that we can physically admit sufficiently short time advances that are within the uncertainty principle, i.e. time delays $\Delta T$ can also get very small negative values:
\begin{equation}
\label{uncertainty}
    \Delta T > -1/\omega.
\end{equation}
This condition is weaker than asymptotic causality (also known as Gao-Wald condition~\cite{gr-qc/0007021}) which demands a positive time delay, i.e.  $\Delta T > 0$.\footnote{This condition is known to imply analyticity of the amplitudes in  local theories~\cite{Toll:1956cya}. However, to the best of our knowledge, it is still an open question whether the polynomial boundedness follows from it, i.e. whether locality and asymptotic causality imply each other.} From a quantum mechanical point of view, the latter might be a too strong requirement, for example it is known to exclude higher derivative interactions in the EFT expansion. Moreover, the weaker condition in Eq.~\eqref{uncertainty} is enough to guarantee the absence of closed time-like curves~\cite{Adams:2006sv,Hollowood:2015elj}. 

A comprehensive study of causality bounds in the case of shift-symmetric scalar theories based on the condition in Eq.~\eqref{uncertainty} was performed in~\cite{CarrilloGonzalez:2022fwg}. We plotted the causality constraints on $\tilde{g}_3$ and $\tilde{g}_4$ derived in~\cite{CarrilloGonzalez:2022fwg} together with our bounds in Fig.~\ref{fig3.b}. We find the following results. (\textit{i})~IR causality bounds are non-compact and significantly weaker than local positivity constraints on the Wilson coefficients, thus leaving a place in the space of EFT parameters for causal EFTs whose UV completions can \textit{only} be non-local. (\textit{ii})~Exponentially bounded amplitudes allow for negative values of $g_4$ that are compatible with IR causality and a non-local UV completion. (\textit{iii})~A causal EFT with cutoff $\Lambda$ can admit a non-local UV completion only if the scale of non-locality is of the same order or larger than the EFT cutoff, i.e.  
\begin{equation}
    \tilde{\alpha}=\alpha\Lambda^4<1.02\quad \Leftrightarrow\quad M\geq (1.02)^{-1/4}\Lambda\simeq \Lambda\,,
\end{equation}
where the second inequality is written in terms of the non-local energy scale $M\equiv 1/\alpha^{1/4}.$ 

It is worth to mention that UV completions whose amplitudes are compatible with exponential boundedness, unitarity, analyticity and IR causality, may in principle still show some causality violation at high energies. Our study cannot say anything about UV causality, but this is an aspect that will be investigated in future work.

However, we have managed to construct exponentially bounded, unitary  amplitudes that satisfy UV causality, in particular the condition of asymptotic causality. An example is given by $A(s,t,u)= C+ \left[F(s)+F(t)+F(u)\right]$, where
\begin{equation}
\begin{aligned}
 F(z)=-\sqrt{\alpha} z \, _2F_2\left(\frac{1}{2},\frac{1}{2};\frac{3}{2},\frac{3}{2};-\alpha z^2\right)+\frac{\sqrt{\pi}}{2} \,\text{erf}\left(\sqrt{\alpha}z\right) \left[\log\left(\sqrt{\alpha}z\right)+i \pi \right],
   \end{aligned}
\end{equation}
and $C$ is some constant. In the Supplemental Material we explicitly show that partial-wave unitarity and asymptotic causality are satisfied by this amplitude.

\textit{\textbf{Conclusions.---}} In this Letter, we derived generalized positivity bounds on the Wilson coefficients of the EFT expansion, assuming that the low-energy theory admits a non-local UV completion according to which the amplitudes are exponentially bounded in the complex $s$ and $t$ planes. 

We showed that the allowed regions in the space of EFT parameters can either overlap or not with those in the local case, depending on the value of the ratio of the non-locality energy scale to EFT cutoff. In particular, we found regions where the corresponding EFTs do not admit a local UV completion, but only a non-local one.
We also compared the new positivity bounds with low-energy causality constraints~\cite{CarrilloGonzalez:2022fwg} and found that causal EFTs can admit non-local UV completions only if the scale of non-locality -- defined as $M=\alpha^{-1/4}$ -- is of the same order or larger than the EFT cutoff $\Lambda$.

To support our main findings, we provided an explicit example that serves as a proof-of-concept for the existence of amplitudes that satisfy all the standard desired properties (i.e. partial-wave unitarity, causality, crossing symmetry) except for the polynomial boundedness.

Understanding non-localizable amplitudes can be relevant for describing field-theoretic aspects of black-hole formation via high-energy scattering~\cite{Dvali:2014ila,Buoninfante:2023dyd}, and for formulating QFT-based UV completions of general relativity, such as infinite-derivative gravity~\cite{BasiBeneito:2022wux} or asymptotically safe gravity~\cite{Knorr:2022dsx}. Indeed, it would also be interesting to compare our positivity bounds with constraints derived in the context of asymptotically safe theories~\cite{Basile:2021krr,Eichhorn:2024rkc,Knorr:2024yiu,Eichhorn:2024wba}. Exploring and classifying different types of viable exponentially bounded amplitudes 
could be an interesting direction for future studies. This has the potential to shed new light on the form of the quantum effective action in UV-complete gravitational theories, as well as on the question of the uniqueness of string theory as the only consistent UV completion of general relativity.

\textit{\textbf{Acknowledgements.---}} The authors are indebted to Ivano Basile, Gia Dvali, Alessia Platania, Xinan Zhou, and Shuang-Yong Zhou for several enlightening conversations, and are grateful to Marianna Carillo-Gonzales for sharing her numerical results on IR causality bounds from Ref.~\cite{CarrilloGonzalez:2022fwg}. A.~T. thanks Alexander Zhiboedov for his hospitality at CERN and numerous deep discussions on the reconstruction of unitary and causal amplitudes. A.~T. thanks GuangZhuo Peng and YoungJun Xu for useful comments and feedback. L.~B. and A.~T. thank The Royal Society for financial support through the International Exchanges Scheme 2022.  L.~B. acknowledges financial support from the European Union’s Horizon 2020 research and innovation programme under the Marie Sklodowska-Curie Actions (grant agreement ID: 101106345-NLQG). The work of A.~T. was supported by the National Natural Science Foundation of China (NSFC) under Grant No. 1234710.

\section*{Supplemental material}

\subsection*{Unitary and causal amplitudes violating Froissart-Martin bound}\label{app:examples}
 
Here we construct working examples of UV complete amplitudes that are exponentially bounded but still satisfy partial-wave unitarity, crossing symmetry and (asymptotic) causality.

\paragraph{{\bf Partial-wave unitarity check.}} 
The unitarity condition on the scattering amplitude formulated in terms of the partial waves reads 
\begin{equation}
    2 \,{\rm Im} f_{l}\left( s\right) \geq \left| f_{l}\left( s\right) \right| ^{2}\,,
\end{equation}
where in four spacetime dimensions $f_l(s)$ can be computed as
\begin{equation}
\label{fl_int}
    f_{l}\left( s\right) =\dfrac{1}{16\pi }\int ^{1}_{-1}{\rm d}x \mathcal{P}_{l}\left( x\right) A\left( s,-\dfrac{s}{2}\left( 1-x\right) ,-\dfrac{s}{2}\left( 1+x\right) \right).
\end{equation}

In this work, we mainly focused on crossing-symmetric amplitudes of the form
\begin{equation}
A(s,t,u)=F(s)+F(t)+F(u)\,.
\end{equation}
Assuming that $F(z)$ can be expanded in series around zero, we can write
\begin{equation}
    F(z)=z^{\beta}\sum_{n=0}^{\infty} a_n z^{2 g n},
\end{equation}
where $\beta$ can also be non-integer. Consequently, we can write an analytic expression of the partial waves for $l>0$ by integrating the latter series term by term, i.e.
\begin{equation}
\label{fl}
    f_l(s)=\dfrac{1}{4\pi }\sum^{\infty }_{n=0}a_n \left( -s\right)^{2n g+\beta}\dfrac{\left[\Gamma \left( \beta+2 n g+1\right) \right] ^{2}}{\Gamma \left( \beta+2ng+l+2\right) \Gamma \left( \beta+2n g  -l+1\right)}.
\end{equation}
For exponentially bounded amplitudes with $F(s)\sim e^{-\alpha s^{2 g}}$ we have $a_n=(-\alpha)^n/n!$.
In this case, from the expression \eqref{fl} it is evident that the partial waves have the same type of essential singularity at infinity ($f_l(s)\sim e^{-\alpha s^{2 g}}$ at large $s$) as that of the full amplitude, that is, they are not polynomially bounded in the complex  $s$-plane. However, the unitarity condition is required only for physical momenta $s\geq 0$, which means that their growth in the complex $s$-plane is not necessarily in contradiction with partial-wave unitarity.

The sum in the Eq.~\eqref{fl} can be expressed in terms of rather complicated hypergeometric functions, but we noticed that it was more efficient to check unitarity by computing partial waves numerically from \eqref{fl_int}. As all partial waves are bounded, the full non-linear unitarity condition can be always achieved by suitably adjusting the overall constant factor. 

The simplest form of an exponentiall bounded amplitude could be 
\begin{equation}
A(s,t,u)\propto e^{-\alpha s^2}+e^{-\alpha t^2}+e^{-\alpha u^2}\,.
\end{equation}
However, it can be shown that this expression does not satisfy partial-wave unitarity for large $s$ due to the fact that the corresponding partial waves heavily oscillate and can give rise to negative imaginary contributions. 
In other words, if we assume
\begin{equation}
{\rm Disc}_s A(s,t,u)\propto e^{-\alpha s^2}+e^{-\alpha t^2}+e^{-\alpha u^2}
\end{equation}
and expand in partial waves, it can be shown that the imaginary parts of the partial waves can be negative for large values of $s$.

\begin{figure}%
    \centering

    \subfloat[\centering ]{\includegraphics[width=0.47\textwidth]{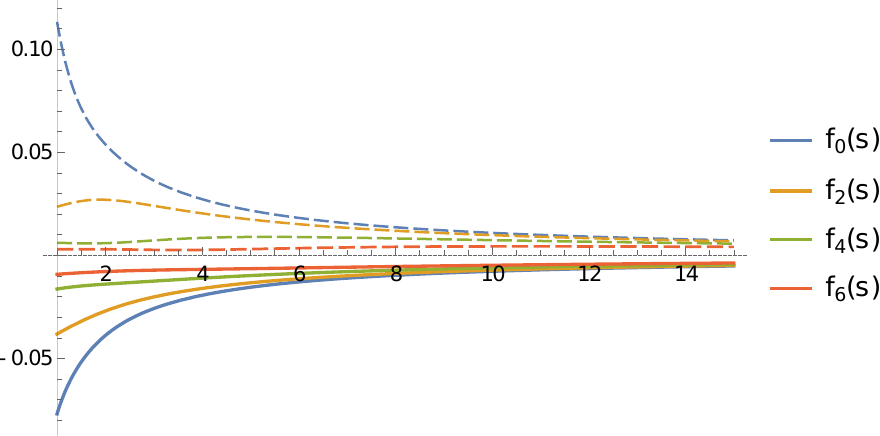}\label{fig4.a}}%
    \qquad
   \subfloat[\centering ]{\includegraphics[width=0.47\textwidth]{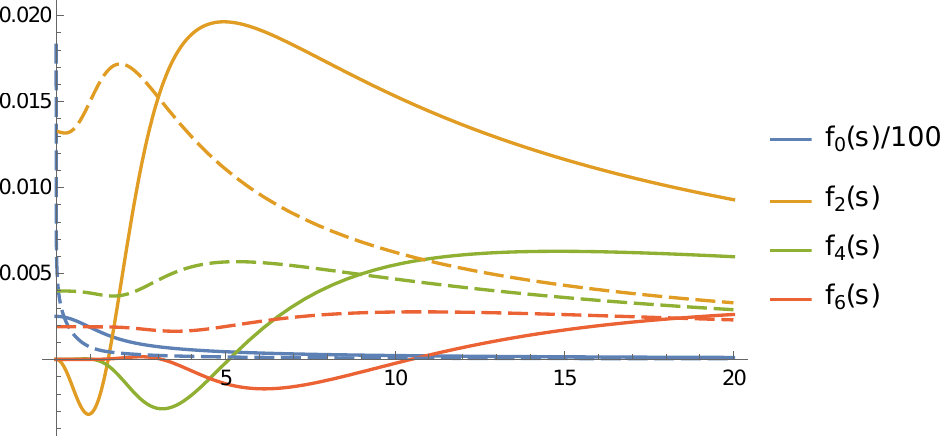}\label{fig4.b}}  
    \caption{Real~(solid lines) and imaginary~(dashed lines) parts of the partial-wave amplitudes as functions of $s$ for the amplitudes~\eqref{alog} (left panel (a)) and~\eqref{Aunitary}  (right panel (b)). We set $\alpha=1$ and $\gamma=1$. } 
    \label{fig:g3h3f3slice}%
\end{figure}

However, we can try to modify the amplitude suitably and check whether we can find expressions that are compatible with unitarity. For example, the amplitude
\begin{equation}
\label{disc-amp-examp}
    {\rm Disc}_s A\propto \left(e^{-\alpha s^2}\log{s}+e^{-\alpha t^2}\log{t}+e^{-\alpha u^2}\log{u}\right),
\end{equation}
has positive-definite $s$-discontinuities of the partial waves. A possible form of the amplitude with this type of discontinuity, but defined in the physical region (i.e. $t\leq 0$), is given by
\begin{equation}
\label{alog}
      A(s,t,u)= C \left[e^{-\alpha s^2}\log{s}\log{(-s)}+
      e^{-\alpha t^2}\log{t}\log{(-t)}+e^{-\alpha u^2}\log{u}\log{(-u)}\right].
\end{equation}
Here the branch cut of the logarithm is specified by the condition $\log{s}-\log{(-s)}=i \pi$. This amplitude can be shown to satisfy partial-wave unitary for reasonable values of the pre-factor $C$. The behavior of the real and imaginary parts of the first partial waves are plotted in Fig. \ref{fig4.a} for $C=1$ and $\alpha=1$.

A weak point of the amplitude~\eqref{alog} is that some of the partial waves blow up in the $s\rightarrow 0$ limit (see Fig. \ref{fig4.a}). Is there a possibility to write an amplitude with finite partial waves in the IR limit? Apparently, it is relatively easy to construct a unitary amplitude that is exponentially bounded. For example, the choice
\begin{equation}
\label{Aunitary}
   F(z)= -i \, \Gamma(\gamma,\alpha z^2) \log{\sqrt{\alpha}z}\,,\qquad \gamma>0,
\end{equation}
where $\Gamma(\gamma, z)$ is the incomplete gamma-function, provides an amplitude whose partial waves have positive and bounded imaginary parts, as shown in Fig.\ref{fig4.b}.

\paragraph{{\bf A probe of causality.}} The notion of causality is typically encoded in an analyticity requirement for the scattering amplitudes~\cite{Toll:1956cya}. However, analyticity itself does not guarantee the absence of superluminal propagation. 
The condition forbidding any time advance in the propagation of a signal (also called asymptotic causality) requires the following inequality to be valid~\cite{2108.05896, hep-th/9709110}:
\begin{equation}
\label{ASC}
    \sqrt{s}\Delta T=2 \sqrt{s} \frac{\partial}{\partial \sqrt{s}}\text{Re}\, \delta_l(s)\approx 2 \sqrt{s} \frac{\partial}{\partial \sqrt{s}}\text{Re} f_l(s)>0\,,  
\end{equation}
Here $\delta_l$ is a scattering phase corresponding to the partial-wave amplitude written as
\begin{equation}
    f_l(s)=-i (e^{i \delta_l(s)} -1)\,,
\end{equation}
and the approximation used means that $f_l(s)$ is small.

\begin{figure}%
    \centering
    \subfloat[\centering ]{\includegraphics[width=0.47\textwidth]{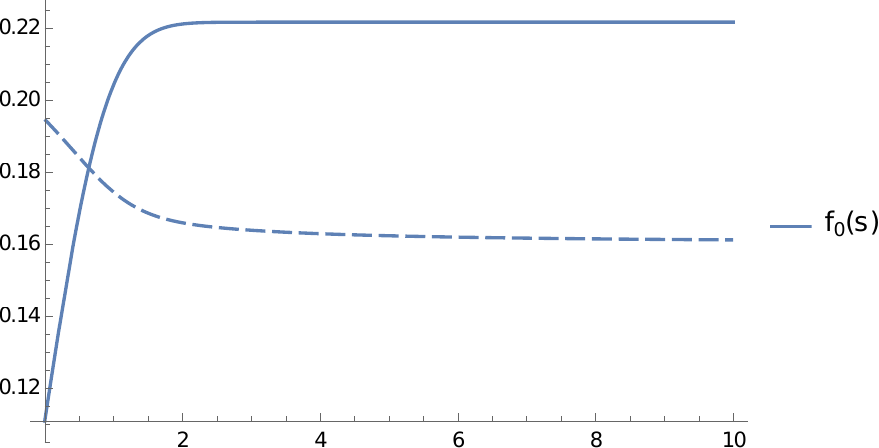}\label{fig5.a}}%
    \qquad
    \subfloat[\centering ]{\includegraphics[width=0.47\textwidth]{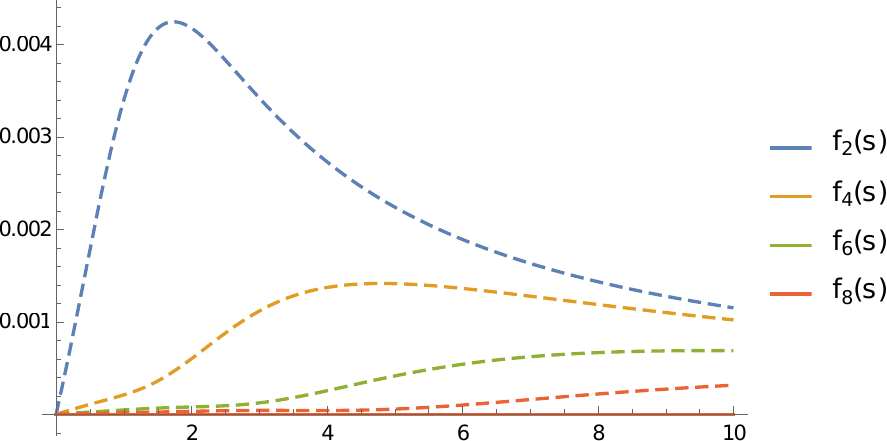} \label{fig5.b}}%
    \caption{(a) Real~(solid lines) and imaginary~(dashed lines) parts of the partial wave $l=0$ for the  amplitude~\eqref{Agood} as a function of $s$. (b) Imaginary parts of the partial waves $l=2,4,6,8$ for the  amplitude~\eqref{Agood} as a function of $s$. We set $C=5 i/2$. All partial waves have positive imaginary parts. The real part of $f_0(s)$ is a monotonically growing function which guarantees that the asymptotic causality condition~\eqref{ASC} is satisfied.} 
    \label{fig:5}%
\end{figure}

Oscillations in the real part of the partial waves can violate the condition of asymptotic causality, indeed this happens for the amplitude~\eqref{Aunitary}. However, we can still find amplitudes that satisfy both partial-wave unitarity and causality. 
An explicit example is by
\begin{equation}
\label{Agood}
     A(s,t,u)= C+ \left[F(s)+F(t)+F(u)\right]\,,
\end{equation}
where
\begin{equation}
\begin{aligned}
  F(z)=&  \int_0^z {\rm d}y \log{\left(-\sqrt{\alpha}y\right)}e^{-\alpha y^2}\\
  =&-\sqrt{\alpha}z \, _2F_2\left(\frac{1}{2},\frac{1}{2};\frac{3}{2},\frac{3}{2};-\alpha z^2\right)+\frac{1}{2} \sqrt{\pi } \,\text{erf}\left(\sqrt{\alpha}z\right) \left[\log
   \left(\sqrt{\alpha}z\right)+i \pi \right],
   \end{aligned}
\end{equation}
and $C$ can be an arbitrary constant. It is important to mention that the presence of a non-zero constant $C$ does not affect any of the dispersion relations used in this paper since the relevant results were obtained by taking derivatives with respect to $s$ and $t$. 

Partial-wave amplitudes corresponding to \eqref{Agood} are plotted in Fig. \ref{fig:5} from which we clearly see that the real parts are monotonically growing functions of $s$ (i.e. there is no oscillation). Note that the only partial wave that has non-zero real part is $l=0.$ The constant term $C$ in \eqref{Agood} is therefore required for the amplitude so that the $l=0$ partial wave satisfies the asymptotic causality. 

This last example provides a proof-of-concept for the existence of amplitudes that are non-localizable (i.e. exponentially bounded) but, at the same time, satisfy partial-wave unitarity and asymptotic causality (i.e. absence of time advances).

Let us remark that the requirements of unitarity and analyticity are not enough to guarantee the validity of asymptotic causality. In fact, the weaker condition of macroscopic causality, which allows for unresolvable time advances could be more justified from a quantum mechanical point of view. The latter is given by
\begin{equation}
   \sqrt{s}\Delta T(s)= 4 s \frac{\partial}{\partial s}\text{Re} f_l(s)> -1.
\end{equation}
We found it much easier to construct examples of amplitudes that satisfy only the macroscopic causality condition. In fact, it suffices to require them to be smooth bounded functions for real and positive values of $s$. In this sense, the unitarity condition implies macroscopic causality, while the asymptotic causality seems to be a much stronger requirement than a combination of analyticity and partial-wave unitarity.

\subsection*{Derivation of the analytic bound on $g_3$}\label{app:analytic-bound-g3}

We provide here more details about the method that was used to derive the bounds involving the coupling $g_3$. 
We refer to~\cite{2012.15849, 2105.02862, 2204.07140} for more detailed derivation and explanations. 
To derive a constraint on the coupling $\tilde{g}_3$, we also need the partial-wave decomposition for the discontinuities in $d=4$ spacetime dimensions, i.e.
\begin{equation}
    {\rm Disc}_s A(s,t)=\sum_{l\,\rm{even}} n_l \rho_l(s)\mathcal{P}_l\left(1+\frac{2t}{s}\right)\,,
\end{equation}
where $n_l=16\pi (2l+1),$ $f_l(s)$ is the partial-wave amplitude, $\rho_l(s)={\rm Im} f_l(s)$ is the spectral density, $\mathcal{P}_l(x)$ are the Legendre polynomials and the angular momentum $l$ only takes even values because we are working with a scalar field theory. 

If we define 
\begin{equation}
    \left[\frac{1}{\mu}\right]\equiv \sum_{l\,\rm{even}} n_l \int_{\Lambda^2}^\infty \frac{{\rm d}\mu}{\pi} \rho_l(\mu) e^{\alpha \mu^2}\frac{1}{\mu}\,,
\end{equation}
take derivatives with respect to $s$ and $t$, and set $s\rightarrow0$, $t\rightarrow0$ afterwards, we can write
\begin{equation}
\begin{split}
    &g_2=\left[\frac{1}{\mu^3}\right]\,,\,\,\, g_3=\left[\frac{3}{\mu^4}\right]-\left[\frac{2L^2}{\mu^4}\right]\,,\,\,\, \\
    &g_4=\left[\frac{1}{2\mu^5}\right]-\left[\frac{\alpha}{2\mu^3}\right].   \label{form:g2g3g4}
    \end{split}
\end{equation}
where $L^2\equiv l(l+1)$. 
Crossing symmetry implies that the relation for the coupling $g_4$ in the EFT expansion can be derived in two different ways. The first way is differentiating four times with respect to $s$ and then taking $s,t\rightarrow0$. The second way is differentiating twice with respect to $s$ and twice with respect to $t$ and then taking $s,t\rightarrow0$. Since the two procedures must be equivalent, we get the so-called \textit{null constraint}:
\begin{equation}
    \left[\frac{L^2(L^2-8)}{\mu^5}\right]=0\,.  \label{form:null}
\end{equation}

If we define
\begin{equation}
    a_{k,q}=\left[\frac{L^{2q}}{\mu^{k+1}}\right]=\sum_{l\,\rm{even}}n_l  \int_{\Lambda^2}^\infty \frac{{\rm d}\mu}{\pi}\rho_l(\mu) e^{\alpha \mu^2}\frac{L^{2q}}{\mu^{k+1}}\,,
    \label{brack-kq}
\end{equation}
%
we can write the following relaxing equalities:
\begin{equation}
    a_{2,0}-a_{3,0}\geq0\,,\quad a_{3,0}-a_{4,0}\geq 0\,,\quad a_{3,1}-a_{4,1}\geq0\,,\dots  \,.\label{form:cons1}
\end{equation}
Calling $f_1(\mu,L^2)=1/\mu^{3/2},\, f_2(\mu,L^2)=1/\mu^{5/2}, \, f_3(\mu,L^2)=L^2/\mu^{5/2},$ and applying the Gram's inequality \cite{20141071}, we obtain
\begin{equation}
     \left|\begin{matrix}
        \int f_1^2 & \int f_1f_2& \int f_1f_3  \\[1.5mm]
        \int f_2f_1 & \int f_2^2& \int f_2f_3 \\[1.5mm]
        \int f_3f_1 & \int f_3f_2& \int f_3^2 
    \end{matrix}\right|\geq0\,\Rightarrow\,
    \left|\begin{matrix}
        a_{2,0} & a_{3,0}& a_{3,1}  \\[1.5mm]
        a_{3,0} & a_{4,0}& a_{4,1} \\[1.5mm]
        a_{3,1} & a_{4,1}& a_{4,2} 
    \end{matrix}\right|\geq0   \,,  \label{form:cons2}
\end{equation}
where for brevity we omitted the measure ${\rm d}\mu$ and the integration range $[\Lambda^2,\infty]$ in the above integrals.
In addition, we also have the rank-2 version of Gram's inequalities,
\begin{equation}
    \left|\begin{matrix}
        a_{2,0} & a_{3,0}  \\[1.5mm]
        a_{3,0} & a_{4,0} \\
    \end{matrix}\right|\geq0\,,~ 
    \left|\begin{matrix}
        a_{2,0} & a_{3,1}  \\[1.5mm]
        a_{3,1} & a_{4,2} 
    \end{matrix}\right|\geq0\,,~
    \left|\begin{matrix}
        a_{4,0} & a_{4,1}  \\[1.5mm]
        a_{4,1} & a_{4,2} 
    \end{matrix}\right| 
    \geq0\,.   \label{form:cons3}
\end{equation}
The final condition follows from the fact that $L^2\in\{0,6,20,42,\dots\}$ is discrete-valued, thus we have 
\begin{equation}
    (L^2-6)(L^2-20)\geq0
\end{equation}
for all possible values of $L^2$ in the discrete set. Thus, we have
\begin{equation}
\begin{split}
  &\left[\frac{(L^2-6)(L^2-20)}{\mu^{5}}\right]\geq 0,~ \\
  &a_{4,2}-26a_{4,1}+120a_{4,0}\geq0.\label{form:cons4}
  \end{split}
\end{equation}

To obtain the constrained regions in the  $a_{3,1}\text{-}a_{4,0}$ plane (corresponding to $g_3\text{-}g_4$ plane), we need to project out $a_{4,2},$ $a_{4,1}$ and $a_{3,0}$. The null constraint \eqref{form:null} can be written as $a_{4,2}=8 a_{4,1}$ and can help us to eliminate $a_{4,2}$. Moreover, to project out $a_{4,1}$ and $a_{3,0}$ it is essential to saturate the most ``loose'' bounds containing $a_{4,1}$. For example, after using the null constraint, Eq.~(\ref{form:cons4}) becomes 
\begin{equation}
    a_{4,0}\geq \frac{20}{3}a_{4,1}\,.
\end{equation}
Thus, we take $a_{4,0}=20a_{4,1}/3$ to eliminate $a_{4,1}$. Consequently, Eq.~({\ref{form:cons2}}) gives
\begin{equation}
    \tilde{a}_{3,1}^2-\frac{40}{3}\tilde{a}_{3,0}\,\tilde{a}_{3,1}+\frac{160}{3}\tilde{a}_{3,0}^2-\frac{80}{9}\tilde{a}_{4,0}\leq0\,, \label{form:rank3}
\end{equation}
where $\tilde{a}_{k,q}\equiv\frac{a_{k,q}}{a_{2,0}}\Lambda^{2(k-2)}$. The last inequality implies
\begin{equation}
    \tilde{a}_{3,1}\leq\frac{20}{3}\tilde{a}_{3,0}+\sqrt{\frac{80}{9}(\tilde{a}_{4,0}-\tilde{a}_{3,0}^2)}\,.
\end{equation}
If we define the function 
\begin{equation}
    f(\tilde{a}_{3,0})\equiv\frac{20}{3}\tilde{a}_{3,0}+\sqrt{\frac{80}{9}(\tilde{a}_{4,0}-\tilde{a}_{3,0}^2)},
\end{equation}
then the most loose bound corresponds the maximum of $f(\tilde{a}_{3,0})$. This maximum value is given by 
$\tilde{a}_{3,0}=\sqrt{\frac{5}{6}\tilde{a}_{4,0}}$. In addition, from the other constraints in Eqs.~\eqref{form:cons1} and~\eqref{form:cons3}, we also have  $\tilde{a}_{4,0}\leq\tilde{a}_{3,0}\leq\sqrt{\tilde{a}_{4,0}}$, which means that $f(\tilde{a}_{3,0})$ can also reach maximum at the boundary of the allowed range for $\tilde{a}_{3,0}$. Therefore, we need to consider these two regions separately,
\begin{equation}\label{form:a-geometry}
\begin{aligned}
    &0<\tilde{a}_{3,1}<\sqrt{\frac{160}{3}\tilde{a}_{4,0}},\qquad\qquad\qquad 0<\tilde{a}_{4,0}\leq\frac{5}{6},  \\
   &\tilde{a}_{3,1}<\sqrt{\frac{80}{9}(\tilde{a}_{4,0}-\tilde{a}_{4,0}^2)}+\frac{20}{3}\tilde{a}_{4,0}, ~~ \frac{5}{6}<\tilde{a}_{4,0}\leq 1. 
   \end{aligned}
\end{equation}
The shape carved by these inequalities is called a-geometry, the result is general and does not rely on the definition of the brackets in Eq.~\eqref{brack-kq}. In our study that involves non-local amplitudes, we need to replace $\tilde{a}_{4,0}=2\tilde{g}_4+\tilde{\alpha}$ and $\tilde{a}_{3,1}=-\frac{1}{2}\tilde{g}_3+\frac{3}{2}\tilde{a}_{3,0}$ into Eq.~\eqref{form:rank3} and then project out $\tilde{a}_{3,0}$ in order to get our final result for the bound on $\tilde{g}_3$.  It is worth mentioning that adding higher-order null constraints would only lead to a very small improvement of the bound in its upper left corner. This does not affect any of our conclusions.


\bibliographystyle{utphys}
\bibliography{References}




\end{document}